# Diagnosing Client Faults Using SVM-based Intelligent Inference from TCP Packet Traces

Chathuranga Widanapathirana Y. Ahmet Şekercioğlu
Paul G. Fitzpatrick Milosh V. Ivanovich Jonathan C. Li
Department of Electrical and Computer Systems Engineering, Monash University, Australia

*Abstract*—**We present the Intelligent Automated Client Diagnostic (IACD) system, which only relies on inference from Transmission Control Protocol (TCP) packet traces for rapid diagnosis of client device problems that cause network performance issues. Using soft-margin Support Vector Machine (SVM) classifiers, the system (i) distinguishes link problems from client problems, and (ii) identifies characteristics unique to client faults to report the root cause of the client device problem. Experimental evaluation demonstrated the capability of the IACD system to distinguish between faulty and healthy links and to diagnose the client faults with 98% accuracy in healthy links. The system can perform fault diagnosis independent of the client's specific TCP implementation, enabling diagnosis capability on diverse range of client computers.**

## I. Introduction

In recent years, technological developments in computer networking have predominantly focused on improving connection media speeds and state-of-the-art applications. In tandem with user demand for high-speed delivery of information, tolerance for performance and connectivity issues has decreased. Due to the complexity and scale of modern communications networks that include a multitude of possible client devices, traditional ″expert knowledge″ or ″rule based″ methods of performance and fault diagnosis are increasingly inefficient and infeasible.

Analysis of packet traces, especially from the Transmission Control Protocol (TCP), is a sophisticated inference based technique used to diagnose complicated network problems in specialized cases. TCP traces contain artifacts related to behavioral characteristics of network elements that a skilled investigator can use to infer the location and root cause of a network fault. The expertise and resources required for trace analysis and inference, however hinders its usability in the conventional fault resolution process of Internet Service Providers (ISPs).

The most common complaint from broadband users is that their ″connection speed is too slow″ [1]. ISPs typically employ experienced technical staff that continuously monitor and resolve performance issues in servers, backbone and access links, but often the true bottleneck of a user's connection speed is actually the client device [2]. Often this is the result of overly conservative default networking parameters supplied with almost all out-of-the-box operating systems. Correct configuration of these parameters with respect to the access network technology can often improve connection speeds and alleviate user dissatisfaction, but in practice these settings are difficult for novice users to manipulate [3]. Thus, many users experience severely degraded network performance even when the networks are underutilized [4]. The Internet2 performance initiative has found that the median bandwidth in their 10 Gb/s backbone in April 2010 was approximately 3.05 Mb/s [5]. Though many solutions have been proposed for improving network traffic conditions, little attention has been given to solving the bottlenecks or diagnosing faults at the end-user.

In this paper, we address the aforementioned issues by introducing a new intelligent inference method for diagnosing network problems using TCP packet traces which we call the Intelligent Automated Client Diagnostic (IACD) system. The system (i) relies only on collection of packet traces upon reporting of a problem, and (ii) focuses on identifying client device faults and misconfigurations.

The rest of this paper is organized as follows. Section II presents the overview of the IACD system. Section III and IV discusses the design of each classifier and the training process. The performance of the system is evaluated in Section V, and Section VI offers a discussion on system characteristics and comparison with other similar work. Conclusions are drawn and future work is discussed in Section VII.

## II. Operational Overview of the Intelligent Automated Client Diagnostics (IACD) system

The proposed IACD system is outlined in Figure 1. The system starts by collecting a TCP packet trace of a known stream of data between the client device and the ISP's access router. This can be initiated by the user through a specially-created web page that activates the trace collection application. This trace is then analysed by the IACD system, which contains two machine learning trained classifiers.

Assuming that the access server is optimized for the connection, performance problems experienced by the end user can be attributed to either the access link or client device. Since the system is designed to diagnose client device problems, a test is first performed to identify whether the performance problem is due to faulty access links.

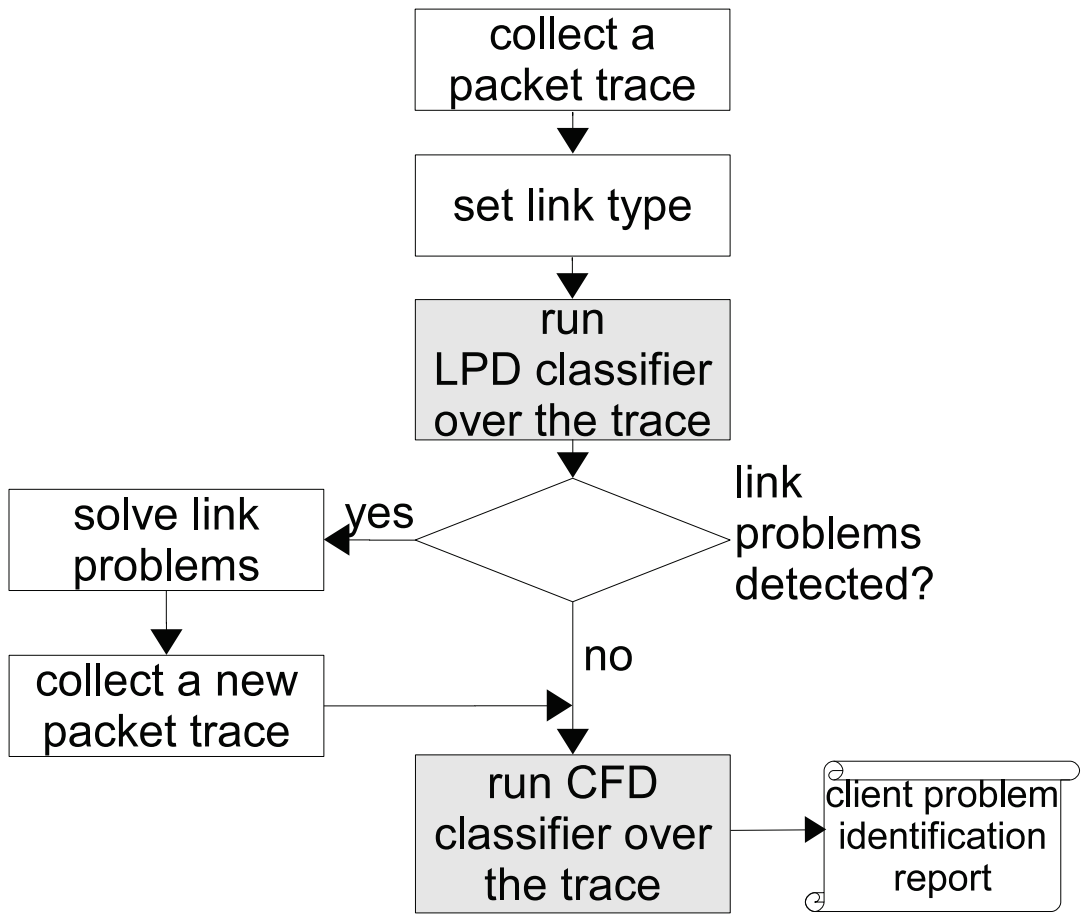

Fig. 1. Overview of the operation of the IACD system

The TCP packet trace is passed to the first stage of the system, the *Link Problem Detection (LPD) classifier* that reports whether an access link is operating as expected. If the LPD classifier determines that the access link is faulty, then the link issues should be resolved before attempting to diagnose any client faults. We have left the automatic diagnosis of the access link problems for future work. If the link is identified to be healthy, the packet trace is passed to the *Client Fault Diagnostic (CFD) classifier* to identify the exact root causes of any problems in the client device.

## III. IACD SYSTEM CLASSIFIERS

### A. Link Problem Detection (LPD) classifier

The LPD classifier detects the artifact patterns which exist only when a link's performance degrades from the expected baseline. We define an access link performing at the expected baseline as a *healthy access link* and a link with degraded performance as a *faulty access link*. The performance expectation of a healthy link is network-specific. The operator has the freedom to train several LPD classifier modules, each trained for a specific link type and baseline performance (e.g. 24 Mb/s or 12 Mb/s DSL link, 14 Mb/s HSDPA link, 54 Mb/s 802.11g link with 1% packet loss or 5% packet loss).

The design, as shown in Figure 2 has two phases: first, the training phase creates an appropriate classifier model using two sets of trace samples from faulty and healthy links. Second, the diagnostic phase uses the trained classifier model to determine the artifacts hidden in an undiagnosed trace.

The training data set $\Theta_{\text{lpd}}$ of $n$ instances is in the form of

$$\Theta_{\text{lpd}} = \{(\mathbf{x}_i, y_i) | \mathbf{x}_i \in \Re^m, y_i \in \{+1, -1\}\}_{i=1}^{n} \quad (1)$$

with $\mathbf{x}_i$ being an $m$-dimensional feature vector and class label $y_i$, either $+1$ for the faulty link or $-1$ for the healthy link, to which each $\mathbf{x}_i$ belongs. For example, a sample trace ($i = 1$) from a faulty link, with four features ($m = 4$) is denoted by $\{0.5, 0.03, 0, 0.99, +1\}_{i=1}$.

We chose the L2 soft-margin SVMs [6], [7] with kernel mapping to model the best non-linear separating hypersurface between the faulty class and the healthy class. For an $m$-dimensional input feature vector, the resultant class boundary is an $m$-dimensional hypersurface that separates the two classes with the maximum margin.

### B. Client Fault Diagnostic (CFD) classifier

The first stage of the IACD system ensures that the access link is not causing the connection problem. The second stage, Client Fault Diagnostic (CFD) classifier identifies the specific types of client faults, if any, that cause the performance problem.

In our CFD classifier design, we opted to use a parallel network of binary classifier modules (CF-classifiers), each structurally similar to LPD classifier in Figure 2 and trained to diagnose a single fault. This arrangement collectively performs a multi-class classification. A network of binary classifiers were chosen over a single multi-class classifier because of the

(i) flexibility to continually add new diagnostic capabilities, without the need to retrain the complete system,
(ii) freedom to select classifier parameters optimized to detect a specific type of artifact independent of other fault classifiers,
(iii) parallelism which can shorten classification time in a scalable manner as the number of modules increases.

The training samples are stored in a trace database

$$\Theta_{\text{cfd}} = \{(\mathbf{x}_i, y_i) \,|\, \mathbf{x}_i \in \Re^m, y_i \in \{\text{cf}_0, \text{cf}_1, ..., \text{cf}_p t\}\}_{i=1}^{n} \quad (2)$$

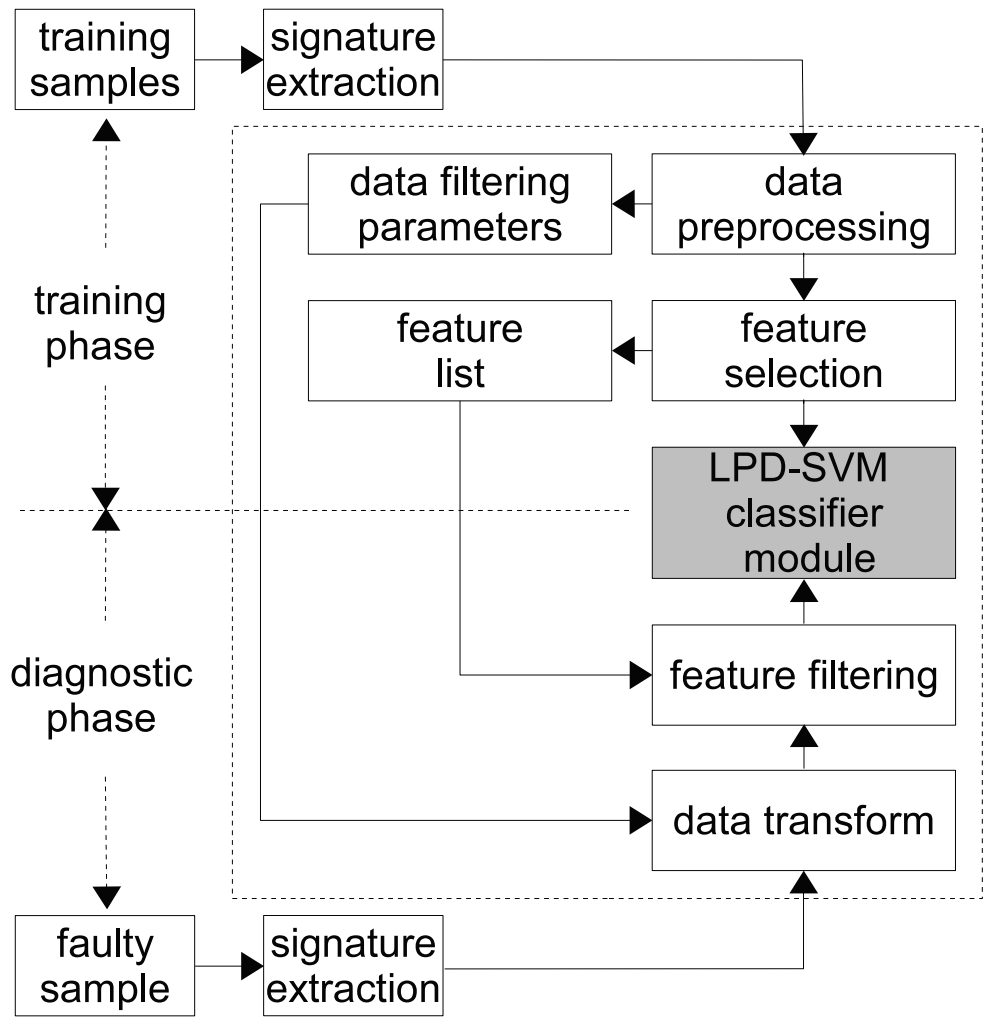

Fig. 2. LPD classifier design.

where $\mathbf{x}_i$ is the $m$-dimensional feature vector and $y_i$ is the class label. The class label $\mathrm{cf}_0$ for a healthy client and $\mathrm{cf}_1, \mathrm{cf}_2, ..., \mathrm{cf}_p$ for $p$ types of different client faults. Each module then selects the training data subset ($\Theta^j_{\mathrm{cf}}$) with traces labeled as $\mathrm{cf}_j$ for the faulty class and $\mathrm{cf}_0$ labeled traces in the healthy class for training the $j^{th}$ binary CF-classifier. Each CF-classifier module in CFD classifier uses a L2 soft-margin SVMs for pattern classification.

## IV. Classifier Training

### A. Data collection

Training samples are collected either from a controlled test bed to emulate the faults in a well-regulated environment or from the actual cloud network. Using standard packet capture libraries, we capture two traces, one at the client and one at the server. Both traces are captured with bi-directional packet flows with file of size 100 MB to ensure most connection details are captured.

### B. Trace signature creation

Two collected packet traces are analyzed and trace statistics are extracted using a tool developed based on *tcptrace* [8] to form an $m$-dimensional feature vector, $\mathbf{x}_i$. The feature vector $\mathbf{x}_i$, combined with the class label $y_i$, is called the *signature* of the $i^{th}$ instance. Our technique extracts 140 different statistical parameters for each trace which forms a combined total of 280 parameters for each signature. The statistical trace characterization technique transforms a packet stream into a data vector encapsulating the connection characteristics and preserving the fault artifacts.

The signatures are unique, even within the same class as shown in Figures 3 and 4. Yet for each type of fault class, there exists a subset of features with common values which are specific for that class. This unique subset of features forms the artifact.

### C. Data pre-processing

The raw feature vectors further processed before being used for classifier training. This step improves the overall classification accuracy by enhancing data coherency and consistency within the classes. First, categorical attributes such as the class labels FAULTY and HEALTHY are converted to numeric data, i.e. +1 for the faulty class and -1 for the healthy class. Then the data is shifted and linearly re-scaled along each feature to fit in the range 0-1. Data re-scaling avoids numerical difficulties and avoids features with greater numeric range dominating the smaller.

### D. Hidden trace artifacts

Figures 3 and 4 show the standardized trace databases, $\Theta_{\mathrm{lpd}}$ and $\Theta_{\mathrm{cfd}}$ used for training the IACD system. The $i^{th}$ row represents the feature vector $\mathbf{x}_i$ of the $i^{th}$ trace sample, mapped to color space for easy visualization of signature characteristics. Null features have been removed for clarity. Figure 3 shows samples from two classes, faulty path ($y_i = -1$) and healthy path ($y_i = +1$) (Equation (1)). Figure 4 contains samples from multiple client fault classes ($\mathrm{cf}_1, ..., \mathrm{cf}_5$) and the healthy client ($\mathrm{cf}_0$) (Equation (2)). Figures 3 and 4 show that the signature extraction process creates unique signatures for every TCP packet trace, even within the same class, preserving the connection characteristics.

In Figure 3, some feature values (columns) behave sporadically (such as features 40-45, 60-65, 160-165 in Figure 3), and provide no usable information to the classifier. However, we can identify a feature subset (features 1-5, 19-22, 115-120, 175-180 in Figure 3) that clearly separates the faulty class from the healthy class. This creates the artifact for this specific case. Using these artifacts as a visual guide, the faulty and healthy access links in Figure 3 can be distinguished.

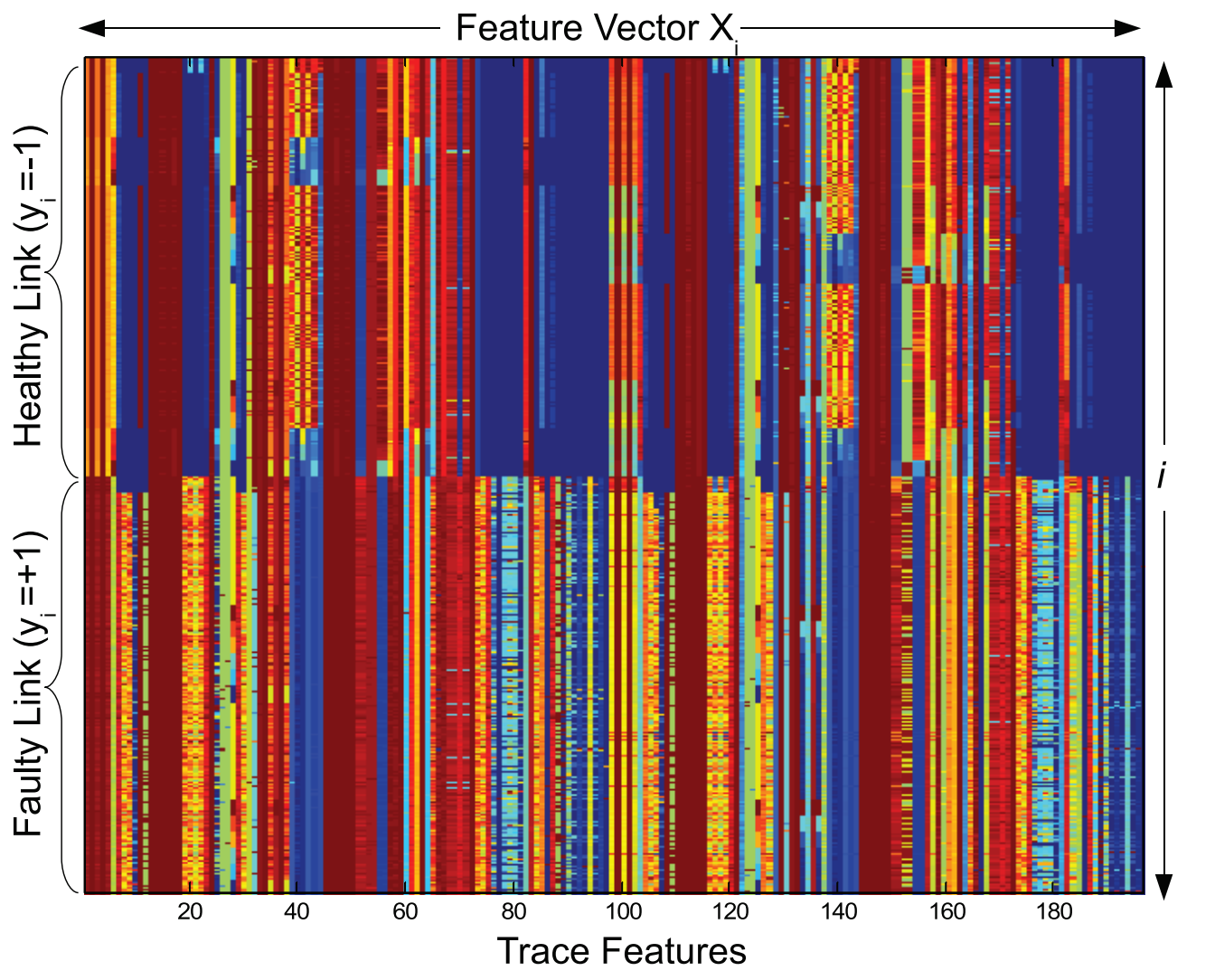

Fig. 3. LPD classifier signature database, $\Theta_{\mathrm{lpd}}$ for comparison of faulty and healthy link trace characteristics.

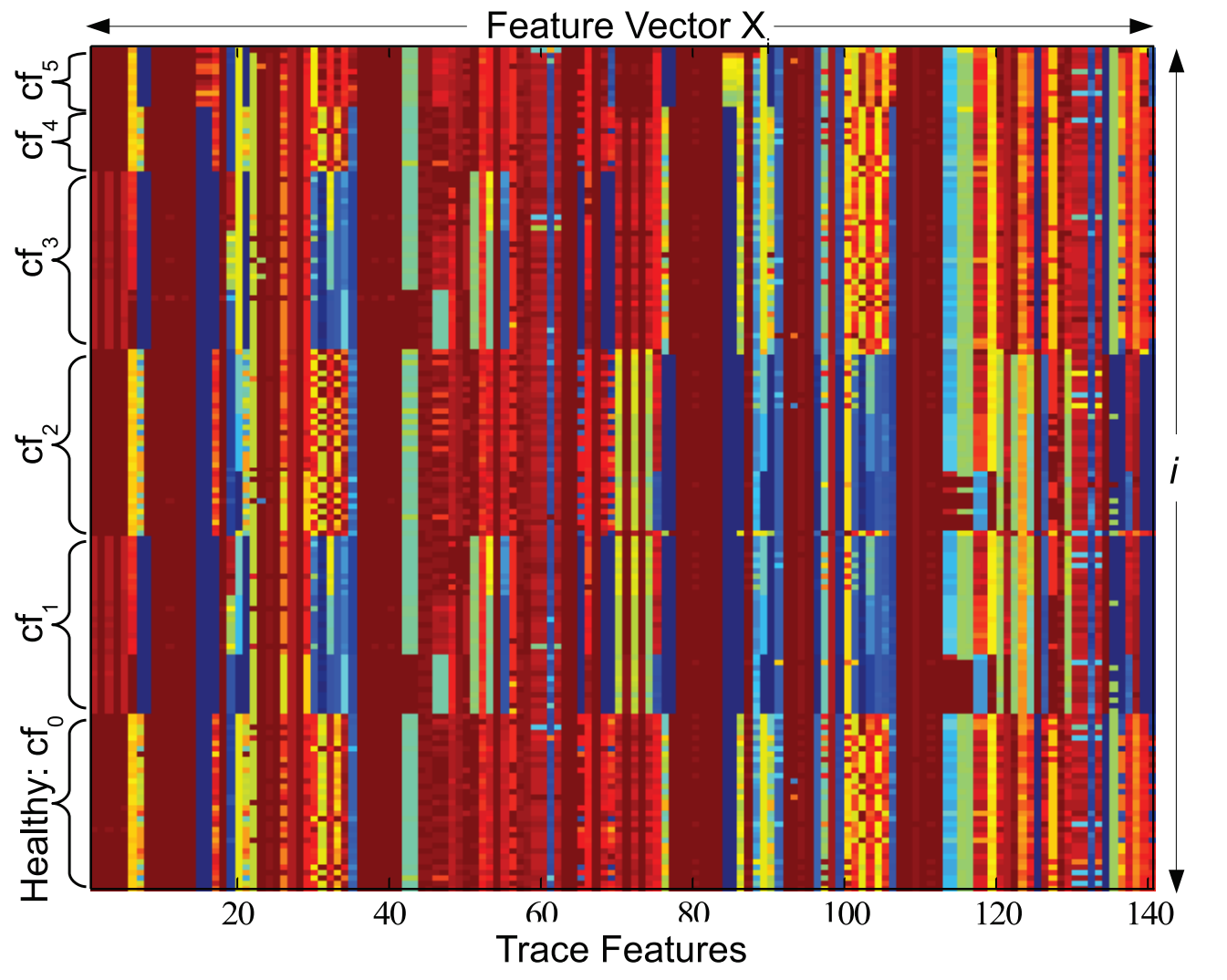

Fig. 4. CFD classifier signature database, $\Theta_{\mathrm{cfd}}$ for comparison of client fault classes, $\mathrm{cf}_i$.

Similarly, Figure 4 shows multiple client fault classes, $cf_i$. The signatures of different fault classes exhibit clear differences compared with the healthy client and are more subtle compared to $\Theta_{\text{lpd}}$. When trained, the IACD system automatically identifies and classifies a trace and produces the visually comprehensible classes shown in Figures 3 and 4.

### E. Feature selection

Although the signature format is identical in every sample, only a particular subset of features contributes to the artifact. We first use a filtering technique, Student's t-test (two sample t-test) to assess the significance of every feature for separating the two classes. Next, the features sorted in the order of significance are cross-validated by incrementing the number of features selected for each class (wrapper technique) against test data to identify the best number of features required for each classifier. The feature selection process reduces the $m$-dimensional feature vector in (2) to $q$-dimensions, where the combination of $q$ features creates the artifact. Note that further fundamental analysis of the relationship between selected features and client faults can be facilitated by, and in turn aid in refining, the feature selection process.

## V. IACD SYSTEM PERFORMANCE

### A. Network emulation

As a proof of concept, data was collected in a network test bed which emulated an access link, client computer and the access server. The client and server ran on Linux 2.6.32 systems (with Ubuntu distribution), capable of running multiple TCP variants. The access link was emulated using a network emulator, dummynet [9] on FreeBSD 7.3. Each box was connected using full-duplex, 1000 Mb/s cat5e ethernet. Different client and link conditions were emulated using the Linux and dummynet parameter configurations.

### B. Experiment Criteria

Experiments included analyzing the performance of the IACD system with one LPD classifier (for a single type of access link), and CFD classifier with a network of four CF-classifier modules. The experimental set-up emulated a full duplex wired access link with a 80 Mb/s bandwidth, 10 ms delay with no packet losses and no packet reordering as the healthy link. Faulty links were emulated by inducing packet losses (from 1% up to 10%) and increased delays (from 15ms up to 100ms). Both the server and the healthy client (Linux 2.6.32) had an protocol stack optimized for the healthy link.

For client faults, we emulated the disabled Selective Acknowledgement (SACK) option (CF-Classifier 1) and the disabled Duplicate Selective Acknowledgement (D-SACK) option (CF-Classifier 2), which have been found to cause performance issues in the high bandwidth connections [10], [11]. Also, Socket buffer limitations, another common and hard to diagnose performance bottleneck [12], [13] were emulated by creating insufficient read buffers (CF-Classifier 3) and write buffers (CF-Classifier 4) at the client as two separate cases. Multiple, simultaneous client faults were emulated by creating both read and write socket buffer limitations at the same time. All buffer limitations were emulated using three buffer levels to collect traces from a range of possible scenarios.

For training data, both the server and client were limited to run TCP-CUBIC, with only 11 traces per each fault class being collected to re-create the worst case practical limitations. To analyze the system performance, we used the following test data sets that collected with

(i) a TCP-CUBIC client
(ii) a TCP-BIC client
(iii) a TCP-NewReno client

in addition to the data set used for training. The data sets (ii) and (iii) were collected to evaluate the TCP agnostic properties of the system with previously unseen TCP variants.

### C. Diagnostic performance of LPD classifier

We used 100 traces for faulty and healthy class, originally with 280 features (before feature selection) for training the LPD classifier. The quadratic kernel was chosen for this particular classifier by cross-validation. Quadratic programming (QP) with a maximum of 1000 iterations was used to solve the optimization problem.

The proposed feature selection technique requires cross-validation before selecting the best feature subset. Although we cross-validated a number of feature subsets, our analysis is limited to two subsets of 75 (Figure 5(a)) and 25 (Figure 5(b)) sorted features. Compared to Figure 3, both of these feature limited databases show a clearer separation between the two classes. The feature

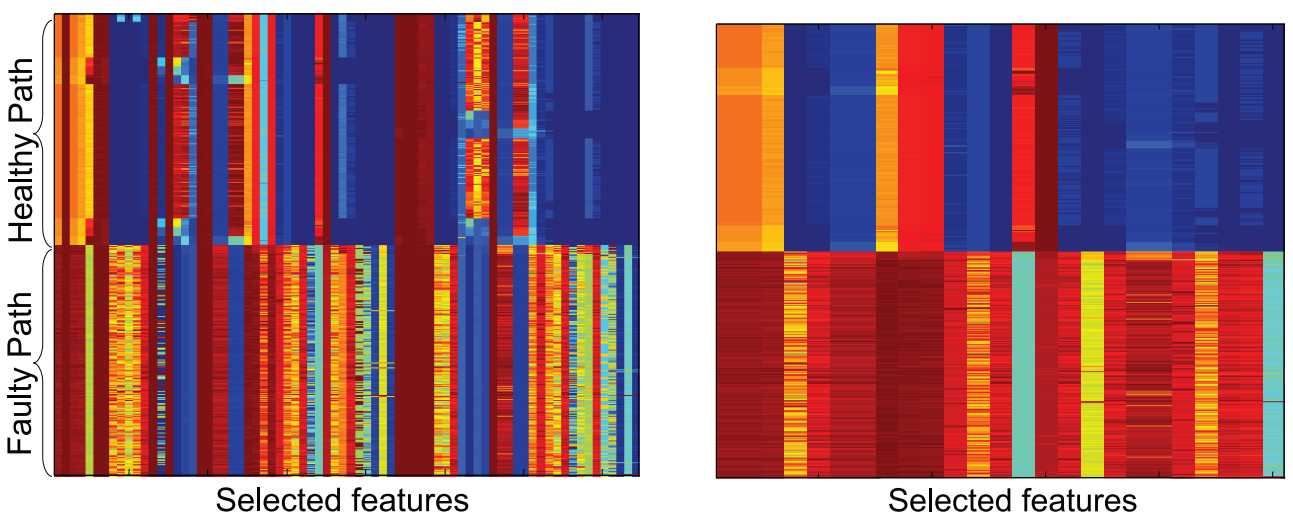

(a) 75 sorted features using Student's t-test feature selection
(b) 25 sorted features using Student's t-test feature selection

Fig. 5. Comparison of trace databases after feature selection.

| Link Problem Detection Accuracies | | |
|---|---|---|
| Trace Samples | 75 features | 25 features |
| TCP-CUBIC training set | 100% | 100% |
| TCP-CUBIC testing set | 100% | 100% |
| TCP-BIC testing set | 99.825% | 100% |
| TCP-NewReno testing set | 97.378% | 100% |

TABLE I
CLASSIFICATION ACCURACY OF LPD CLASSIFIER FOR DETECTING FAULTY LINKS.

selection technique has reduced the dimensionality of the problem by 73% and 91% with 75 and 25 feature subsets, respectively.

Each of the four testing data sets consisted of 264 traces (132 faulty, 132 healthy link), collected with healthy and faulty clients. The 75 features used for training the classifier create a more complicated class boundary compared to 25 features, a phenomenon called boundary over-fitting. With over-fitted boundaries, even a small deviation of the data (vector) at the boundary can cause a misclassification. For the first test using the training instances, the classification accuracies were 100%. The accuracy remained at 100% during the second test when the previously unseen TCP-CUBIC data set was used. In these two cases, the lack of outlier samples resulted in high classification accuracy, even with an over-fitted boundary in the case of 75 features. However, the over-fitted boundaries resulted in classification errors of 0.175% for TCP-BIC, and 2.622% for TCP-NewReno when the artifacts subtly deviated from those of TCP-CUBIC. When the dimensionality was reduced to 25 features, the LPD classifier created a more generalized boundary capable of compensating for artifact variations. As a result, the classifier was highly successful in separating the two classes with 100% accuracy for both the TCP-BIC and TCP-NewReno cases.

### D. *Diagnostic performance of CFD classifier*

The choice of classifier module parameters has a significant impact on the performance. Table II shows the parameters chosen for each of the four CF-classifiers. The cross-validation technique for selecting the parameters and feature subset considered not only the individual classifier accuracy, but also the possible false positives.

The Table III shows the diagnostic accuracy of the CFD classifier, which considers the collective output of the CF-classifier network. When tested with the CUBIC training and testing data sets, the system was capable of diagnosing the client's disabled SACK option, disabled D-SACK option, read buffer limitation and write buffer limitations with high accuracy. Similarly, when tested with TCP-BIC and TCP-NewReno, variants not used during the training phase, the four client faults were diagnosed with 100% accuracy. These results demonstrated the TCP-independent nature of the proposed CFD classifier design.

| Non-linear SVM Parameters | | |
|---|---|---|
| | SACK problem | DSACK problem |
| Kernel | Linear | RBF |
| Features | 12 | 32 |
| | Read Buffer | Write Buffer |
| Kernel | $3^{rd}$ degree polynomial | RBF |
| Features | 24 | 16 |

TABLE II
SVM PARAMETERS USED IN EACH CF-CLASSIFIER MODULE OF THE CFD CLASSIFIER.

| Client Fault Diagnostic Accuracies | | | |
|---|---|---|---|
| Trace Samples | SACK | DSACK | WBuff |
| CUBIC training | 100% | 100% | 100% |
| CUBIC testing | 100% | 100% | 93.94% |
| BIC testing | 100% | 100% | 100% |
| New-Reno testing | 100% | 100% | 100% |
| | | | |
| Trace Samples | RBuff | R-WBuff | Healthy |
| CUBIC training | 100% | | 94.81% |
| CUBIC testing | 96.36% | 96.97% | 93.50% |
| BIC testing | 100% | 96.90% | 92.10% |
| New-Reno testing | 100% | 100% | 91.71% |

TABLE III
DIAGNOSTIC ACCURACY OF THE CFD CLASSIFIER, DERIVED FROM THE COLLECTIVE OUTPUT OF THE CF-CLASSIFIER NETWORK.

The healthy clients were identified with a 94.81% and 93.5% accuracies during the first two tests of TCP-CUBIC training and test data sets. When samples from healthy clients with TCP-BIC and TCP-NewReno were tested, the detection accuracies were at 92.10% and 91.71%, marginally lower than the other cases. This is due to the slightly higher tendency of obtaining a false positive in at least one of the CF-classifiers by healthy clients' traces compared to other samples. When presented with traces taken from clients with simultaneous read and write buffer deficiencies, CF-classifier 3 and CF-classifier 4 were capable of independently identifying the faults from the trace. This capability led to a collective diagnostic accuracy of 96.97%, 96.90% and 100% for CUBIC, BIC and NewReno data sets, respectively.

## VI. SYSTEM CHARACTERISTICS AND COMPARISON WITH THE SIMILAR WORK

For the root cause diagnosis of client performance problems, the proposed IACD system offers many advantages over the other available trace inference methods:

- The system offers a fully-automated, comprehensive framework which is extendible to diagnose a diverse range of faults, contrary to the limited capabilities of offered by tools that uses TCP traces for information gathering and measurement purposes [14]–[16].
- Diagnostic capability of the system evolves with the diversity of the fault signature databases, instead of the inference algorithm. Users can collaborate to create common signature repositories, encompassing a wide range of faults, networks, and client platforms. Most rule based systems are limited to a specific set of faults and lack the generality to operate effectively in a dynamic environment [17]–[19].

- The system relies solely on packet traces collected at end-points and can be implemented as an application. This provides flexibility for the operator to deploy the IACD system at any desired network location. Popular client diagnostic solutions, mainly based on Web100 TCP kernel instrumentation require changes to the kernel and the system itself [17], [20].
- End-user systems can be diagnosed without remotely accessing or physically logging on to the systems; a capability unavailable in many network diagnostic tools. Most machine learning based solutions such as *NEVERMIND* [21], *pinpoint* [22], *Netprints* [16] require information such as user requests, event logs, system calls or private network traffic which demands privileged access.
- The proposed technique, contrary to many other similar work, avoids both the idiosyncrasies of individual TCP implementation and the usage of TCP flags as an information source [14], [19]. Instead, the connections are characterized using per-connection statistics without relying on the negotiated flags and independent of the TCP variant.

## VII. Conclusion

In this work, we present the IACD system, and evaluated its performance under controlled conditions. The results show that the LPD classifier can effectively identify and separate out the link problems without being affected by the client behavior and TCP type. Also, with a small number of training samples, CFD classifier produces high diagnostic accuracy.

To our knowledge, the IACD system is the first framework for automating the client diagnosis with TCP packet trace-based fault signatures and SVM-based learning. This work provides the foundation to extend the system to more complex, real world network environments with thousands of users, diverse client platforms, and complex traffic patterns.